\documentclass[11pt,twoside]{article}
\usepackage{asp2010}

\resetcounters

\begin{document}

\title{Abundance Patterns in S-type AGB stars : Setting Constraints on Nucleosynthesis and 
Stellar Evolution Models}
\author{Pieter Neyskens$^1$, Sophie Van Eck$^1$, Bertrand Plez$^2$, St\'{e}phane Goriely$^1$,\\
Lionel Siess$^1$ and Alain Jorissen$^1$
\affil{$^1$Institut d'Astronomie et d'Astrophysique, Universit\'{e} Libre de Bruxelles,\\
CP 226, Boulevard du Triomphe, B-1050 Bruxelles, Belgium}
\affil{$^2$GRAAL, Universit\'{e} Montpellier-II, CNRS-UMR 5024,\\
Place Eug\`{e}ne Bataillon, F-34095 Montpellier, France}
}

\begin{abstract}
During the evolution on the AGB, S-type stars are the first objects to experience s-process nucleosynthesis 
and third dredge-ups, and therefore to exhibit s-process signatures in their atmospheres. 
Their significant mass loss rates (10$^{-7}$ to 10$^{-6}$ M$_{\odot}$/year) make them major contributors 
to the AGB nucleosynthesis yields at solar metallicity. Precise abundance determinations 
in S stars are of the utmost importance for constraining e.g. the third dredge-up luminosity 
and efficiency (which has been only crudely parameterized in all current nucleosynthetic models so far). 
Here, dedicated S-star model atmospheres are used to determine precise abundances of key 
s-process elements, and to set constraints on nucleosynthesis 
and stellar evolution models. A special interest is paid to technetium, an element with no stable isotopes 
($^{99}$Tc, the only isotope produced by the s-process in AGB stars, has a half-life of $2.1\times 10^{5}$ years). 
Its detection is considered as the best signature that the star effectively populates 
the thermally-pulsing AGB phase of evolution. The derived Tc/Zr abundances are compared, as a function of 
the derived [Zr/Fe] overabundances, with AGB stellar model predictions. The [Zr/Fe] overabundances are 
in good agreement with the model predictions, while the Tc/Zr abundances are slightly overpredicted. This 
discrepancy can help to set better constraints on nucleosynthesis and stellar evolution models of AGB stars. 
\end{abstract}

\section{Introduction : S-type stars and the role of technetium}
Based on their spectra, late-type stars are classified in three main groups: M, S and C stars. These 
groups differ in their surface chemical composition. 
Spectra of M stars are characterized by strong TiO absorption bands, while absorption bands of C$_{2}$ and other carbon-rich
molecules appear in C star spectra. 
Stars showing absorption bands of ZrO (in addition to TiO absorption bands)
are classified as S. The presence of ZrO presumes that S stars are located on the thermally-pulsing 
asymptotic giant branch (TPAGB), where the repeated occurrence of thermal pulses and third dredge-ups (3DUP) 
permits carbon and s-elements to be synthesized and brought to the stellar surface \citep{2001NuPhA.688...45V}. 
Therefore S stars are believed to be transition objects between oxygen-rich M and carbon-rich 
C stars \citep{1983ARA&A..21..271I}.\\
The presence of technetium (Tc, Z=43, an element with no stable isotope) in some S-star spectra
\citep{1952ApJ...116...21M} 
has led to a dichotomy among the S stars: $\it{intrinsic}$ S stars exhibit Tc lines in their 
spectrum, while $\it{extrinsic}$ S stars lack Tc absorption lines. The laboratory half-life of $^{99}$Tc (the isotope
produced by the s-process in AGB stars) is 2.1$\times$10$^{5}$ years, which is of the order of the time spent
by low- and intermediate-mass stars on the TPAGB. This implies that $\it{intrinsic}$ S stars are genuine TPAGB
stars in contrast to $\it{extrinsic}$ S stars. $\it{Extrinsic}$ S stars did not produce their heavy elements themselves, but 
accreted them in the past from a close-by, now extinguished, companion TPAGB star \citep{1999A&A...345..127V}. 
Hence, $\it{extrinsic}$ S stars are low-luminosity giant stars belonging to a binary system
\citep{1999A&A...345..127V,2000A&A...360..196V}.\\
Today, one of the major uncertainties in the AGB stellar evolution models has to deal with the s-process neutron source. 
Observations and models hint $^{13}$C($\alpha$,n)$^{16}$O, instead of $^{22}$Ne($\alpha$,n)$^{25}$Mg, 
as being the dominant neutron source 
in low-mass AGB stars \citep{1986ApJ...311..843S, 2000A&A...362..599G}.
The efficiency of the $^{13}$C neutron source depends on the amount of protons mixed 
down from the H-rich envelope into the C-rich intershell [Partial Mixing, PM, which triggers the chain 
$^{12}$C(p,$\gamma$)$^{13}$N($\beta$)$^{13}$C].
Large uncertainties exist on the depth and on the formation of the $^{13}$C pocket. 
Different mechanisms like convective overshooting, rotation and gravity waves 
have been proposed \citep{1996A&A...313..497F, 2004A&A...415.1089S, 2003MNRAS.340..722D}. 
Details about the 3DUP are also badly known due to our lack of knowledge about convective mixing 
mechanisms in (AGB) stars. Furthermore, AGB modelling uncertainties do also arise from reaction rates uncertainties.\\
A detailed abundance analysis of Tc, Zr and Fe is able to set better constraints on the large uncertainties of the 
mixing and nucleosynthetic processes inside AGB stars since different assumptions on the PM, 
the s-process nucleosynthesis and the 3DUP will lead to different abundance predictions of carbon and s-elements 
\citep{2000A&A...362..599G}.
The surface enrichment in Tc is, in contrast to other s-elements, hampered by its 
radioactive decay. This property gives rise to an almost time-independent surface abundance of Tc during 
the ascent of the AGB, while the abundance of all the other s-elements increases progressively 
\citep{2001NuPhA.688...45V}.
However, it is a hard task to derive accurate chemical abundances in AGB stars, especially in S stars.
The thermal structure of S stars is, in contrast to M stars, also influenced by the non-solar C/O ratio and 
s-process abundances. The use of equivalent widths to derive abundances of individual elements in S stars 
involves large errors due to the strong molecular blending, 
which makes it hard to access the continuum flux.
A higher precision in abundance determinations could be obtained by spectrum synthesis from 
a S-star model atmosphere. 

S-star \rm{MARCS} model atmospheres were computed with
the latest version of the \rm{MARCS} code for late-type stars \citep{2008A&A...486..951G}. 
The S-star model atmospheres are covering a large set of atmospheric parameters (see Van Eck et al. this volume). 
The resulting synthetic spectra are compared with high-resolution observed spectra
to obtain abundances of Tc, Zr and Fe for 7 $\it{intrinsic}$ S stars from the sample of 205 Henize S 
stars \citep{1960AJ.....65..491H} (This sample contains all $\it{intrinsic}$ and $\it{extrinsic}$ 
S stars with R $\leqslant$ 10.5 and $\delta$ $\leqslant$ -25$^{\circ}$).
\section{Observational data for the Henize S stars and S-star $\rm{MARCS}$ model atmospheres}
Optical low-resolution Boller $\&$ Chivens spectra ($\Delta\lambda$=3~\AA , 4400~\AA-8200~\AA) were obtained for a large fraction of the Henize S stars, in January 1997 at La Silla with the 1.52m ESO telescope.
High-resolution Coud\'{e} Echelle spectra ($\Delta\lambda\approx$0.08~\AA, 4220~\AA-4280~\AA) were also 
taken around that time with the 1.4m CAT telescope at ESO.
The high-resolution spectra are covering two Tc I lines: Tc I 4238.191~\AA~and Tc I 4262.270~\AA. This implies that the presence
or absence of Tc can be checked in a consistent way. Optical Geneva 
UBV photometry and infrared SAAO JHKL photometry (Catchpole, priv. comm.) are also available \citep{2000A&AS..145...51V}.\\
The extended grid of \rm{MARCS} model atmospheres for S stars contains 3522 models for 2700~K $\leq$ T$_{\rm eff}$ $\leq$ 4000~K (steps 
of 100~K), 0.0 $\leq$ log g $\leq$ 5.0 (steps of 1.0), [Fe/H]=0.0 and -0.5, C/O = 0.501, 0.751, 0.899, 0.925, 0.951,
0.971 and 0.991, and [s/Fe] = 0.0, +1.0 and +2.0 dex. These 1D models are constructed 
with the assumptions of homogeneous stationary layers, hydrostatic and local 
thermodynamic equilibrium, and energy conservation by radiative and convective fluxes. The opacity sampling
technique is used and different continuous and line opacities are taken into account.
Properties of the \rm{MARCS} model atmospheres for S stars and the resulting synthetic spectra are discussed elsewhere 
(Gustafsson et al. 2008; Van Eck et al. this volume).
\articlefigure[height=7.5cm,width=10cm]{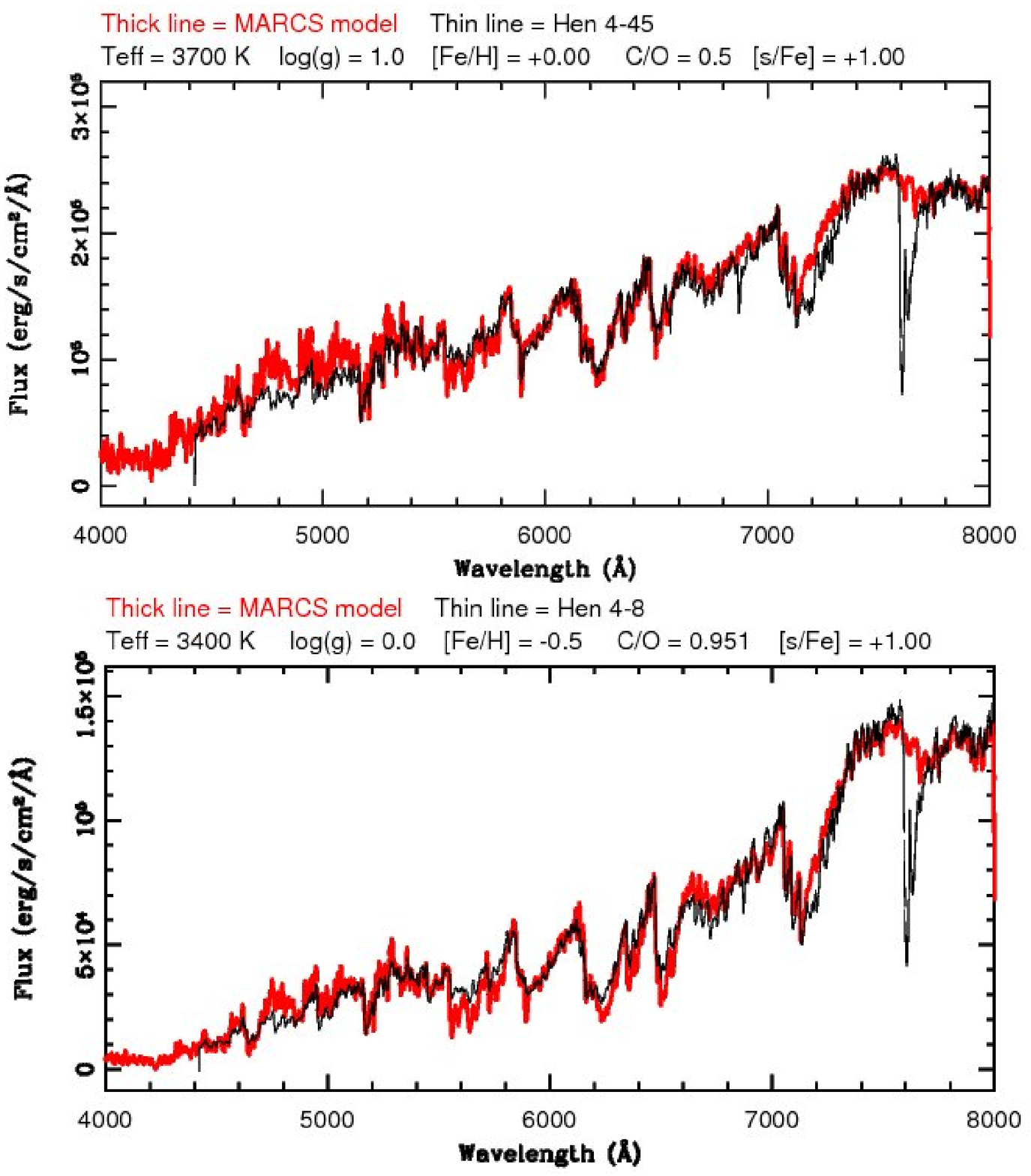}{fig1}{A comparison between the observed low-resolution optical spectrum 
(thin black line) and the selected synthetic spectrum (thick red line) for Henize 4-45 (top) and 
Henize 4-8 (bottom). The atmospheric parameters of the synthetic spectra are given at the top of each panel. The 
7600~\AA~feature is from telluric O$_{2}$.}
\section{Deriving abundances of intrinsic S stars}
\articlefigure[height=3.1cm,width=11.5cm]{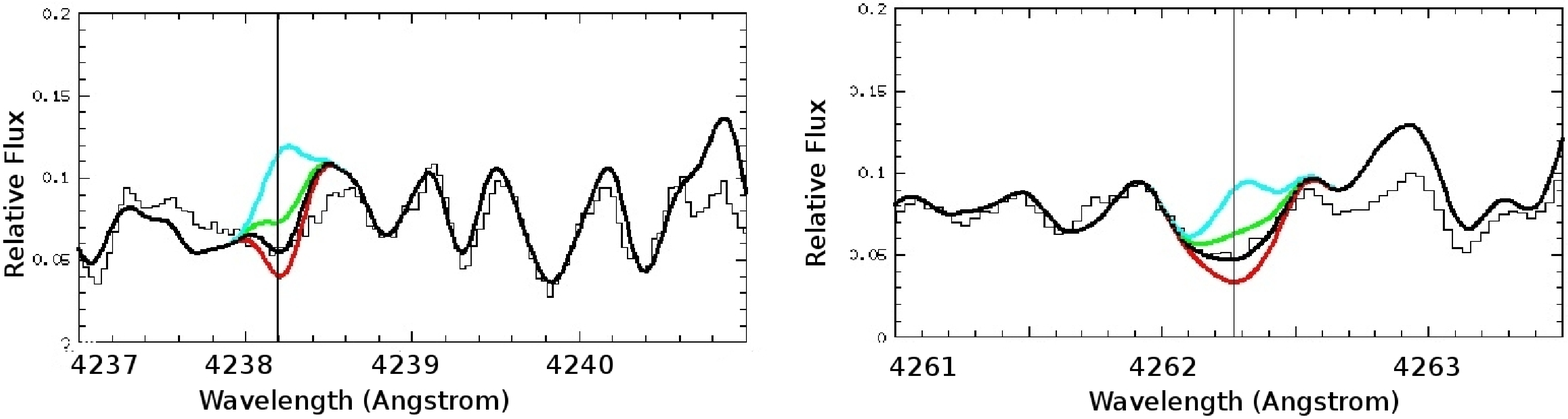}{fig3}{A synthesis of the two Tc lines (Tc I 4238.191~\AA~and
Tc I 4262.270~\AA) for 4 different Tc abundances (No Tc, log~$\varepsilon$(Tc)~=~0.0, 
log~$\varepsilon$(Tc)~=~0.3 and log~$\varepsilon$(Tc)~=~0.6) compared with the high-resolution observation (thin line) of Hen 4-162.}
\articlefigure[height=7.2cm,width=10cm]{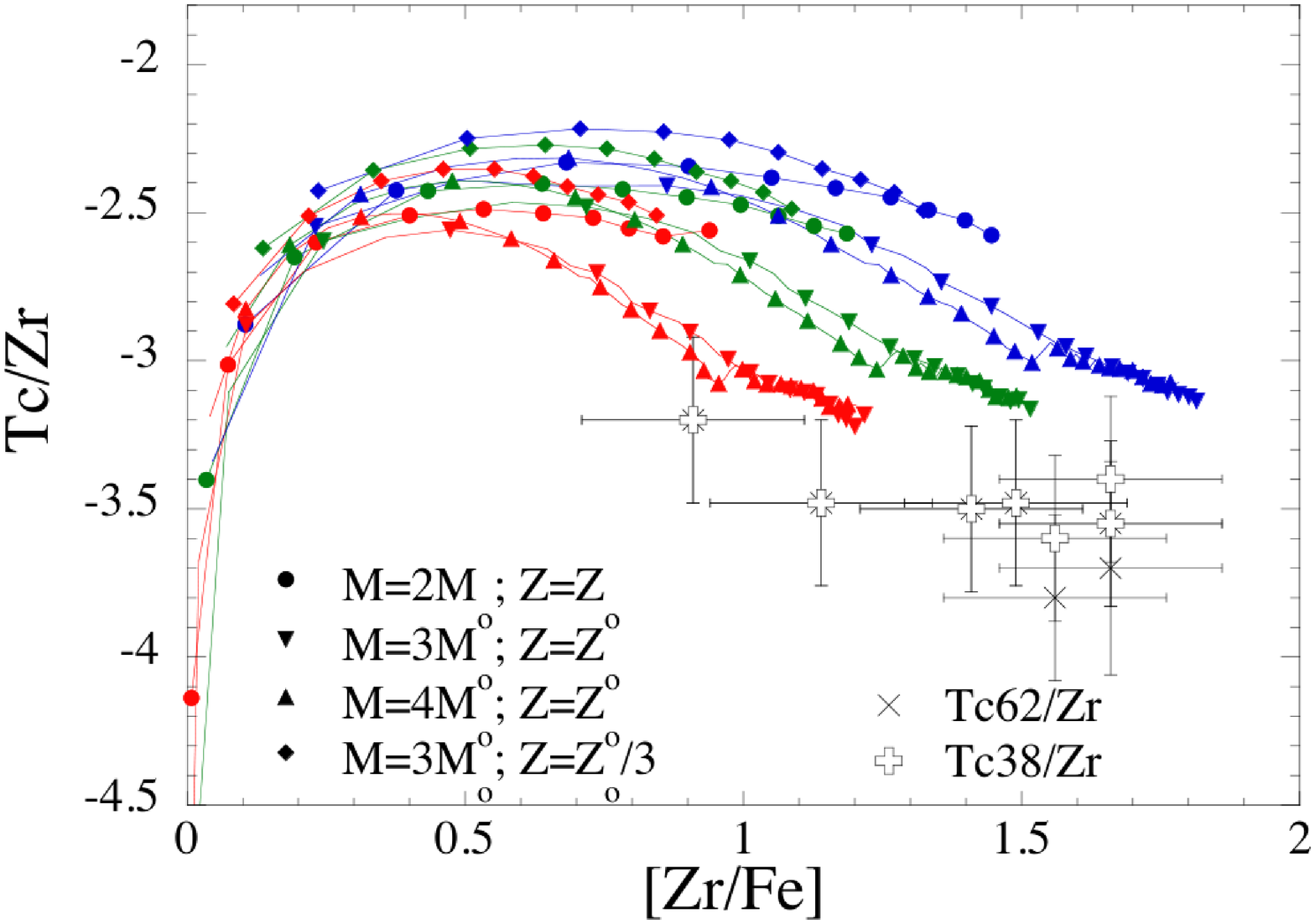}{fig2}{Comparison between observed and predicted Tc/Zr ratios 
as a function of the [Zr/Fe] overabundances. Four model stars with different masses and metallicities 
are considered (M~=~2, 3, 4~M$_{\odot}$ at solar metallicity and M~=~3~M$_{\odot}$ for 
Z=Z$_{\odot}/3$). For each star, the envelope s-process enrichment is calculated all along 
the AGB phase (the symbols  correspond to the abundance ratio just after the third dredge-up) 
assuming 3 different values for the extent  of the partial mixing zone $\lambda_{pm}$: 
for each star (i.e given symbols), the leftmost curve corresponds to $\lambda_{pm}$=0.05, 
the rightmost to $\lambda_{pm}$=0.20 and the middle one to $\lambda_{pm}$=0.10. 
Observations are shown by the two cross types. Tc62 and Tc38 are the Tc abundances derived from the Tc I 4262.270~\AA~and 
Tc I 4238.191~\AA~lines respectively.}
To derive abundances in S stars, from the comparison between optical high-resolution observed spectra 
and synthetic spectra, one has  first to find the atmospheric parameters. 
The estimate of the stellar parameters is based on the comparison between 
observed and synthetic colors (UBV $\&$ JHKL) and between observed and synthetic band-strengths (TiO, ZrO and NaD). 
More information about this so-called ``best-model finding tool'' is given in Sects.~3 and 4 
of Van Eck et al.~(this volume).
The results are shown in Fig.~\ref{fig1}.
This figure compares (for two different Henize stars) dereddened observed 
low-resolution spectrum with the synthetic spectrum derived from the selected ``best'' S-star \rm{MARCS} model.
The dereddening of the observed spectra is based on \citet{1989ApJ...345..245C}. 
Besided photometry, these visual comparisons serve as checks for the derived atmospheric parameters, 
since the stellar parameters largely influence the spectral shape and the strength of the TiO and ZrO bands.\\
Synthetic high-resolution spectra, based on the derived atmospheric parameters, are then compared with high-resolution
observed spectra. This allows us to fine-tune the individual chemical abundances to obtain a 
good match between the observed and synthetic spectra. In this way, abundances of Tc, Zr and Fe were obtained.
Fig.~\ref{fig3} shows a fit of the two Tc I lines at 4238.191 \AA~and 4262.270 \AA~respectively.
\section{Comparison between the derived abundances and the stellar evolution models}
The final Tc/Zr abundances are shown in Fig.~\ref{fig2} as a function of the [Zr/Fe] overabundances. 
The observations are also compared in Fig.~\ref{fig2} with post-processing
s-process calculations performed on 4 AGB models computed with the stellar evolution code
STAREVOL \citep{2007A&A...476..893S}. The s-process model considered here 
corresponds to the partial mixing of protons into the C-rich region at the time of the third dredge-up, 
as described in detail in \citet{2000A&A...362..599G}. The extent of the partial mixing zone is varied 
between 5\% and 20\% of the extent of the thermal pulse ($\lambda_{pm}$ corresponds to the 
ratio of the mass of the partial mixing zone to the mass of the thermal pulse at its maximum extension), 
so that the s-process enrichment in the stellar envelope reaches values compatible with the observations, 
i.e about 1 to 2 dex enrichment for [Zr/Fe]. As can be seen in Fig.~\ref{fig2}, the Tc/Zr ratio 
is systematically overpredicted. Different explanations for this discrepancy can be given, 
in particular the partial decay of Tc, either during longer interpulse phases (e.g  in lower-mass stars) 
or during hotter thermal pulses (at T=3~$\times$~10$^{8}$~K, the $^{99}$Tc half-life is reduced from 
2~$\times$~10$^{5}$~yr to 
about 9~yr). These new observations (complemented with Nb abundance determination) can provide strong constraints 
regarding the s-process in AGB stars. The present data will be further analyzed and interpreted in a forthcoming paper.

The derivation of S-star atmospheric parameters (T$_{\rm eff}$, log g, [Fe/H], C/O and [s/Fe]) and individual abundances, 
and the comparison with stellar evolution AGB models, will be done on a larger sample of S stars in the near 
future. The agreement between infrared synthetic spectra (constructed from the derived atmospheric parameters) and 
infrared observed spectra of S stars will also be tested.  
\acknowledgements P.N. is $\it{Boursier}$ $\it{F.R.I.A.}$, Belgium. S.V.E., S.G. and L.S. are F.R.S.-F.N.R.S. research associates. 
This research has been supported by the \it{Communit}\'{e} \it{Fran\c{c}aise} de Belgique~-~\it{Actions} 
\it{de} \it{Recherche} \it{Concert\'{e}es}.
\bibliographystyle{asp2010}
\rm \bibliography{neyskens.bib}
\end{document}